\documentclass[amsmath,amssymb,aps,prd,twocolumn,superscriptaddress,showpacs]{revtex4}
\usepackage{multirow}
\usepackage[dvips]{graphicx}
\usepackage{times}
\usepackage{color}

\newcommand{\Comments}[1]{}
\newcommand{\nn}{\nonumber}

\newcommand{\VEV}[1]{\langle{#1}\rangle}
\newcommand{\chibar}{{\bar{\chi}}}
\newcommand{\Psfig}[2]{\includegraphics[width=#1]{Figs/#2}}
\newcommand{\Feff}[2]{{\cal F}_\mathrm{#1}^{#2}}

\newcommand{\Fq}{{\cal F}_\mathrm{q}}

\newcommand{\Vp}{V^{+}}
\newcommand{\Vm}{V^{-}}
\newcommand{\bsig}{b_\sigma}
\newcommand{\bsigp}{b'_\sigma}

\newcommand{\mq}{m_q}
\newcommand{\tilmq}{\tilde{m}_q}

\newcommand{\Fint}{{\cal D}}
\newcommand{\Seff}[1]{S_\mathrm{eff}^{(#1)}}
\newcommand{\comment}[1]{}

\newcommand{\Zchi}{Z_{\chi}}
\newcommand{\Zp}{Z_{+}}
\newcommand{\Zm}{Z_{-}}

\newcommand{\tilmu}{\tilde{\mu}}
\newcommand{\tilY}{\widetilde{Y}}
\newcommand{\tilD}[1]{\widetilde{D}_{#1}}

\newcommand{\psiT}{\psi_{\tau}}
\newcommand{\psiS}{\psi_s}
\newcommand{\psiTT}{\psi_{\tau\tau}}
\newcommand{\psiTS}{\psi_{\tau s}}
\newcommand{\psiSS}{\psi_{ss}}
\newcommand{\psibarT}{\bar{\psi}_{\tau}}
\newcommand{\psibarS}{\bar{\psi}_s}
\newcommand{\psibarTT}{\bar{\psi}_{\tau\tau}}
\newcommand{\psibarTS}{\bar{\psi}_{\tau s}}
\newcommand{\psibarSS}{\bar{\psi}_{ss}}

\newcommand{\bt}{\beta_\tau}

\newcommand{\btt}{\beta_{\tau\tau}}

\newcommand{\bp}{\beta_{p}}
\newcommand{\Pbar}{\bar{P}}
\newcommand{\lbar}{\bar{\ell}}
\newcommand{\Lbar}{\bar{L}}
\newcommand{\bfx}{{\bf x}}

\newcommand{\etabar}{\bar{\eta}}
\begin{document}
%\preprint{}

\title{
Chiral and deconfinement transitions
in strong coupling lattice QCD\\
with finite coupling and Polyakov loop effects
}

\author{Takashi Z. Nakano}
\email[]{tnakano@yukawa.kyoto-u.ac.jp}
%\thanks{}
\affiliation{Department of Physics, Faculty of Science, Kyoto University,
Kyoto 606-8502, Japan}
\affiliation{Yukawa Institute for Theoretical Physics, Kyoto University,
Kyoto 606-8502, Japan}
%%%%%
\author{Kohtaroh Miura}
\affiliation{INFN-Laboratori Nazionali di Frascati, I-00044, Frascati(RM), Italy}
%%%%%
%%%%%
\author{Akira Ohnishi}
\affiliation{Yukawa Institute for Theoretical Physics, Kyoto University,
Kyoto 606-8502, Japan}

\date{\today}

\begin{abstract}
We investigate chiral and deconfinement transitions
in the framework of the strong coupling lattice QCD for color SU(3)
with one species of unrooted staggered fermion
at finite temperature and quark chemical potential.
We take account of the leading order 
Polyakov loop terms as well as
the next-to-next-to-leading order ($1/g^4$) fermionic terms
of the strong coupling expansion in the effective action.
We investigate the Polyakov loop effects
by comparing two approximation schemes,
a Haar measure method (no fluctuation from the mean field)
and a Weiss mean-field method (with fluctuations).
The effective potential is obtained 
in both cases, and we analytically clarify
the Polyakov loop contributions to the effective potential.
The Polyakov loop is found to suppress the chiral condensate
and to reduce the chiral transition temperature at $\mu=0$,
and 
the 
chiral transition temperature roughly reproduces the Monte Carlo results
at $\beta=2N_c/g^2 \lesssim 4$.
The deconfinement transition is found to be 
the crossover and first order for light
($a m_0 \lesssim 4$ at $\beta=4$) and heavy quark masses, respectively.
\end{abstract}
\maketitle
\section{Introduction}
The phase transition in Quantum Chromodynamics (QCD) 
at finite temperature ($T$) and/or quark chemical potential ($\mu$)
is attracting much attention in recent years.
When thermodynamic parameters such as $T$ reach to a typical QCD scale,
abundantly formed hadrons start to overlap with each other
and matter with color charges would appear,
thus a quark-gluon plasma (QGP)~\cite{Muller:2006ee}
is expected to emerge at such high $T$ and/or $\mu$.
The first principle studies based on the lattice QCD Monte-Carlo (MC) simulations
actually indicate a phase transition from hadron phase to QGP
around $T\simeq 160$-$190$ (MeV)~\cite{Tchid}.
Various experimental observations and theoretical arguments imply
the formation of a strongly coupled QGP in heavy-ion collisions at RHIC.
This extreme form of matter will be extensively investigated
in LHC-ALICE experiments at CERN,
where we can simulate the matter evolution in the early universe.
We also expect the transition to quark matter
when the baryon chemical potential $N_c\mu$ overcomes the nucleon mass energy
$N_c\mu > M_N$.
The color superconductor  
could appear at high density
due to the attraction between quarks in the color antisymmetric
channel~\cite{Alford:2007xm},
and confined dense quark (quarkyonic) matter 
is also expected from large $N_c$ arguments \cite{QY_McLerran}.
The forthcoming experiments at FAIR, J-PARC and
the low energy programs at RHIC
would be informative for these intermediate and high density states of matter,
which may be realized in the neutron-star core.

The QCD phase transition has two aspects;
the chiral and deconfinement transitions.
These transitions are associated with the spontaneous breaking of
the global symmetry $SU(N_f)_L\times SU(N_f)_R$ in the chiral limit
and the global $Z_{N_c}$ symmetry in the heavy quark mass limit,
respectively,
where the order parameters are the chiral condensate and the Polyakov loop.
Theoretical frameworks describing the dynamics of these order parameters
would be mandatory to understand the QCD phase transition and phase diagram.
The lattice QCD MC simulation
is the most reliable and rigorous framework
to investigate the non-perturbative aspects of QCD~\cite{Karsch:2001cy}.
Recent MC simulations demonstrate
that the two transitions occur simultaneously within the error bars
at small or zero chemical potential~\cite{Tchidsame}.
There is no clear explanation for the mechanism of coincidence of transitions,
and some models suggest that the chiral and deconfinement transitions
can start to separate as quark chemical potential becomes large.
Unfortunately, MC simulations have the notorious sign problem
at finite $\mu$~\cite{MC-sign-problem},
and cannot answer whether two transitions can be separated or not.
As a result, we need to invoke some approximations in QCD
or effective models at finite chemical potentials,
such as
the strong coupling lattice QCD~\cite{MDP,Munster:1980iv,DKS,Faldt:1985ec,DHK1985,Fukushima:2003vi,SC-LQCD-phase-diagram,Nishida:2003fb,Azcoiti:2003eb,deForcrand:2009dh,Ilgenfritz:1984ff,Gocksch:1984yk,FukushimaGOmodel,V_Ploop,NLOSC-LQCD,Nakano:2009bf,Bilic,Jolicoeur:1983tz,Kawamoto:1981hw,Langelage,KlubergStern:1982bs,PNLO},
Nambu-Jona-Lasinio model (NJL)~\cite{Hatsuda:1994pi} and Polyakov loop extended NJL model (PNJL)~\cite{Fukushima:2003fw,Ratti:2005jh},
and chiral random matrix model~\cite{Shuryak:1992pi}.

Strong coupling lattice QCD (SC-LQCD)
can provide a simple and lattice-based description
for the chiral and deconfinement transitions at finite $T$ and $\mu$.
This method is based on the strong coupling $(1/g^2)$ expansion,
where the link variables are integrated out analytically
and it is possible to elucidate important aspects of the QCD phase transitions
beyond a contamination of the sign problem.
%SC-LQCD has been shown to describe the nature of the confinement
%and spontaneous breaking of chiral symmetry in the vacuum at strong coupling.
In the pure Yang-Mills theory, 
the color confinement in vacuum was shown in the strong coupling limit
via the area law for the Wilson loop~\cite{Wilson:1974sk}.
SC-LQCD of the pure Yang-Mills theory with higher order corrections~\cite{Munster:1980iv}
is recently extended to the deconfinement transition study at finite $T$~\cite{Langelage}.
At finite $T$, the deconfinement transition was qualitatively explained
on the base of the mean-field treatment of the Polyakov loop
in the leading order of the strong coupling expansion~\cite{Kogut:1981ez},
and higher order corrections on the Polyakov loop action has been
investigated recently~\cite{Langelage}.
The deconfinement transition is also investigated by taking account of
the Haar measure contributions to the effective potential~\cite{V_Ploop}.
The behavior of the Polyakov loop is investigated also in the weak-coupling regime~\cite{Brambilla:2010xn}.

We find some pioneering works on
the chiral symmetry breaking in vacuum~\cite{Kawamoto:1981hw}
and its restoration at finite $T$ and/or $\mu$~\cite{DKS,Faldt:1985ec,DHK1985}
where one can find a sophisticated expression of
the effective potential describing the chiral transition
in the strong coupling limit~\cite{DKS,Faldt:1985ec,DHK1985}.
It is remarkable that a promising phase diagram structure has been obtained
even in the strong coupling limit (SCL)
analytically~\cite{SC-LQCD-phase-diagram,Fukushima:2003vi,Nishida:2003fb,Azcoiti:2003eb}
and in MC simulation \cite{deForcrand:2009dh}.
In our previous works~\cite{NLOSC-LQCD,Nakano:2009bf},
we have developed a formulation to evaluate the next-to-leading order 
(NLO, ${\cal O}(1/g^2)$)~\cite{Faldt:1985ec,Bilic,NLOSC-LQCD},
and the next-to-next-to-leading order 
(NNLO, $({\cal O}(1/g^4)$)~\cite{Jolicoeur:1983tz,Nakano:2009bf}
of the strong coupling expansion.
Finite coupling effects appear as modifications of the quark mass,
quark chemical potential, and the wave function renormalization factor.
The phase diagram evolution with $\beta=2N_c/g^2$ 
can be interpreted in terms of those modifications.
The SC-LQCD technique developed in these works is also applied
to the graphene system in the condensed-matter physics~\cite{Araki:2010gj},
where the strong coupling gauge theory is applicable at low energies.

In the works mentioned above, either of the two order parameters is included,
and the interplay between these order parameters is not discussed.
One of the interesting developments
can be found in the works
by Gocksch and Ogilvie~\cite{Gocksch:1984yk}
and by Ilgenfritz and Kripfganz~\cite{Ilgenfritz:1984ff}.
They developed a model including both of
the chiral condensate ($\sigma$) and Polyakov loop ($\ell$).
In this model, abbreviated as the GOIK model,
the deconfinement is governed by
the effective action for $\ell$ derived
in the leading order of the strong coupling expansion
in the Pure Yang-Mills theory,
and the the chiral transition is described
by the strong coupling limit effective action for quarks
~\cite{Gocksch:1984yk,Ilgenfritz:1984ff,FukushimaGOmodel}.
The coupling between $\sigma$ and $\ell$
is found to appear naturally from the quark determinant.
This coupling leads to the correlation of the chiral condensate 
and Polyakov loop, and the two transitions take place
at similar temperatures~\cite{FukushimaGOmodel}.
This understanding in the GOIK model
led to the development of the Polyakov loop extended Nambu-Jona-Lasinio (PNJL)
model~\cite{Fukushima:2003fw}.

%%%%%%%%%%%%%%%%%%%%%%%%%%%%%%%%%%%%%%%%%%%%%%%%%%%%%%%%%%%%%%%%%%%%%%%%%%%%%%%%
\begin{table*}[hbt]
\caption{
Classification of SC-LQCD at finite $T$ and $\mu$.
The unrooted staggered fermion is utilized in the quark effective action
except for \cite{Langelage},
where the Wilson fermion is included in the hopping parameter expansion.
In Ref.~\cite{Kogut:1981ez}, they treat the chiral spin model, 
which is essentially the same as that of the Polyakov loop action. 
}
\label{Table:DoubleSCE}
\begin{tabular}{c|c|c|c}
\hline
\hline
~& \multicolumn{3}{|c}{Polyakov loop action} \\
\cline{2-4}
Chiral quark action
	& w/o Polyakov loop & Leading Order & Higher orders \\ \hline
w/o quarks
	& ---
	& \parbox[c]{4cm}{
	  ~\\[-0.3ex]
		Kogut, Snow, Stone~\protect\cite{Kogut:1981ez}\\
  		Polonyi, K.~Szlachanyi; Gross;
  		Bartholomew et al.
		~\protect\cite{V_Ploop}\\
	  ~\\[-0.8ex]
	}
	& Langelage, M\"unster, Philipsen~\protect\cite{Langelage}
\\ \hline
LO (SCL, ${\cal O}(1/g^0)$)
	& \parbox[c]{4cm}{
	  "SCL-LQCD"\\
	  Refs.~\protect{\cite{Kawamoto:1981hw,DKS,Faldt:1985ec,DHK1985,SC-LQCD-phase-diagram,Fukushima:2003vi,Nishida:2003fb,Azcoiti:2003eb}}
	}
	& \parbox[c]{4cm}{
	  ~\\[-0.3ex]
	    "GOIK model"\\
	    Gocksch, Ogilvie~\protect{\cite{Gocksch:1984yk}}\\
	    Ilgenfritz, Kripfganz~\protect{\cite{Ilgenfritz:1984ff}}\\
	    Fukushima~\protect{\cite{FukushimaGOmodel}}\\
	  ~\\[-0.8ex]
	}
	& ---
\\ \hline
NLO~(${\cal O}(1/g^2)$)
	& \parbox{4cm}{
	  ~\\[-0.3ex]
	  "NLO SC-LQCD"\\
	  Faldt, Petersson~\protect\cite{Faldt:1985ec}\\
	  Bilic et al.~\protect\cite{Bilic}\\
	  Miura, Nakano, Ohnishi, \\
	  Kawamoto~\protect\cite{NLOSC-LQCD}\\
	  ~\\[-0.8ex]
	}
	& \multirow{2}{4cm}{\parbox{4cm}{
	  "P-SC-LQCD"\\
	  Nakano, Miura, Ohnishi
	  (present work)\\
	  Miura, Nakano, Ohnishi~\protect{\cite{PNLO}}
	}}
	& ---
\\ \cline{1-2} \cline{4-4}
NNLO~(${\cal O}(1/g^4)$)
	& \parbox{4cm}{
	  ~\\[-0.3ex]
	  "NNLO SC-LQCD"\\
  	  Jolicoeur et al.~\protect{\cite{Jolicoeur:1983tz}}\\
	  Nakano, Miura, Ohnishi~\protect{\cite{Nakano:2009bf}}\\
	  ~\\[-0.8ex]
	}
	&
		& ---
\\
\hline
\hline
\end{tabular}
\end{table*}
%%%%%%%%%%%%%%%%%%%%%%%%%%%%%%%%%%%%%%%%%%%%%%%%%%%%%%%%%%%%%%%%%%%%%%%%%%%%%%%%

From a view point of the rigorous strong coupling expansion,
there is a problem to define the expansion order
when we include both of the chiral condensate and Polyakov loop;
the quark effective action has the strong coupling limit ${\cal O}(1/g^0)$,
while the leading order Polyakov loop action
is proportional to $1/g^{2N_\tau}$, where $N_\tau$ is the temporal lattice size.
We therefore should regard the order of the strong coupling expansion
separately in the quark and Polyakov loop effective action.
Namely we start from the leading order in each of the quark
and Polyakov loop action, ${\cal O}(1/g^0)$ and ${\cal O}(1/g^{2N_\tau})$, respectively,
and extend the framework by including higher order terms
of the strong coupling expansion in two directions.
As shown in Table~\ref{Table:DoubleSCE},
the effective action adopted in the GOIK model is the starting point,
where both of the quark and Polyakov loop effective action terms are
in the leading order.
In this effective action, the QCD coupling constant appears only in the coefficient
of the Polyakov loop action,
and that coefficient $J = (1/g^2N_c)^{N_\tau}$ is
replaced with an expression $J \simeq \exp(-\kappa a^2/T)$ using the string tension 
$\kappa$~\cite{Ilgenfritz:1984ff,Gocksch:1984yk,FukushimaGOmodel}.
This is not fully consistent with the quark effective action
where the coupling constant is assumed to be infinite.
Therefore it would be more favorable to combine the quark effective action
in SC-LQCD with finite coupling 
effects~\cite{Faldt:1985ec,Bilic,NLOSC-LQCD,Jolicoeur:1983tz,Nakano:2009bf}
with the Polyakov loop action.
Since the NLO and NNLO effects in SC-LQCD are represented
as modifications 
in the form of the SCL effective action~\cite{NLOSC-LQCD,Nakano:2009bf},
we can combine them with the Polyakov loop effects
in a similar way to that in the GOIK model.
This
{\em Polyakov loop extended strong coupling lattice QCD (P-SC-LQCD)}
would be a promising framework to investigate 
the relation between the chiral and deconfinement transitions,
and has some advantages over other effective models.
For example, P-SC-LQCD is directly based on the lattice QCD,
and we can compare its results with those in MC simulations
at zero chemical potential.

In this paper,
we develop a P-SC-LQCD framework by combining
the leading order Polyakov loop effective action
and NNLO quark effective action.
We derive an analytic expression of the effective potential
in P-SC-LQCD at finite $T$ and $\mu$,
and investigate the chiral and deconfinement phase transitions at $\mu=0$.
The whole phase diagram structure including 
finite $\mu$ is soon reported elsewhere~\cite{PNLO}.
Here we adopt one species of unrooted staggered fermion
which corresponds to four flavors in the continuum limit.
The unrooted staggered fermion is suitable to developing
an analytic formulation,
since its simple structure enables us to obtain the effective potential
analytically.
The effective action is composed of
the fermionic and pure gluonic parts.
We adopt the fermionic effective action with
finite coupling effects~\cite{Faldt:1985ec,Bilic,NLOSC-LQCD,Jolicoeur:1983tz,Nakano:2009bf}.
The leading order Polyakov loop action is obtained 
from 
$N_\tau$ plaquette configuration
as shown in Fig.~\ref{Fig:Polyakov-loop-SC-LQCD}.
We evaluate the Polyakov loop effects and derive the effective potential
in two kinds of methods.
One is the Haar measure method
where we replace the Polyakov loop with its mean-field value.
Instead of integrating out the temporal link variables,
the logarithm of the Haar measure is included in the effective potential.
The other is the Weiss mean-field method,
which includes the fluctuation effects of the Polyakov loop.
In this method, we bosonize the Polyakov loop action and
carry out the temporal link integral.
On the basis of these effective potentials,
we investigate the chiral and deconfinement phase transitions simultaneously.
We also compare the results of the two methods.

This paper is organized as follows.
In Sec.~\ref{sec:Feff}, we derive the effective potential
with Polyakov loop and finite coupling effects 
in both Haar measure and Weiss mean-field methods.
In Sec.~\ref{sec:chiraldeconf},
we investigate the chiral and deconfinement phase transitions,
and compare the results in the two methods.
In Sec.~\ref{sec:CR}, we summarize our work and give a future perspective.

Throughout this paper, we utilize the lattice unit $a=1$,
and physical values are represented as dimensionless values normalized
by the lattice spacing $a$.

%%%%%%%%%%%%%%%%%%%%%%%%%%%%%%%%%%%%%%%%%%%%%%%%%%%%%%%%%%%%%%%%%%%%%%%%%%%%%%%%
%%%%%%%%%%%%%%%%%%%%%%%%%%%%%%%%%%%%%%%%%%%%%%%%%%%%%%%%%%%%%%%%%%%%%%%%%%%%%%%%
%%%%%%%%%%%%%%%%%%%%%%%%%%%%%%%%%%%%%%%%%%%%%%%%%%%%%%%%%%%%%%%%%%%%%%%%%%%%%%%%

\section{Effective Potential}\label{sec:Feff}

\subsection{Lattice QCD action}\label{subsec:QCDaction}
In the lattice QCD, the partition function and action
with one species of unrooted staggered fermion
for color $\mathit{SU}(N_c)$ 
in the Euclidean spacetime
are given as,
\begin{align}
\label{Eq:ZLQCD}
{\cal Z}_{\mathrm{LQCD}} 
=& \int \Fint[\chi,\chibar,U_\nu]~e^{-S_\mathrm{LQCD}}
\ ,\\
S_\mathrm{LQCD}=&S_F+S_G
\ ,\\
S_F
=&\frac {1}{2} \sum_x \sum_{\rho=0}^d 
  \left[ \eta_{\rho,x} \bar{\chi}_x U_{\rho,x} \chi_{x + \hat{\rho}} 
  - \eta_{\rho,x}^{-1} \bar{\chi}_{x + \hat{\rho}}
   U_{\rho,x}^{\dagger} \chi_x \right]
\nn\\
+&m_0 \sum_x \bar{\chi}_x \chi_x 
\ ,\\
S_G
 =&
  \frac {2N_c}{g^2} \sum_{P} \left[
 	1 - \displaystyle \frac{1}{2N_c} \left( U_P + U_P^\dagger \right) \right] ,
\end{align}
where
$\chi (\bar{\chi})$, $m_0$, $U_{\rho,x}$,
and
$U_P$ 
denote
the quark (antiquark) field,
the bare quark mass, link variable,
and plaquette, respectively.
In the staggered fermion,
the spinor index does not appear explicitly and 
its structure is expressed in the staggered phase factor 
$\eta_{\rho,x}=(\eta_{0,x},\eta_{j,x})=(e^\mu, (-1)^{x_0+\cdots +x_{j-1}})$~\cite{Susskind:1976jm,Sharatchandra:1981si,Kawamoto:1981hw}.
The quark chemical potential $\mu$ on the lattice
is introduced as a weight of the temporal hopping
in the staggered factor~\cite{Hasenfratz:1983ba}.
The above lattice QCD action has a global symmetry $U(1)_L \times U(1)_R$ 
in the chiral limit ($m_0\to 0$),
and a staggered chiral transformation is given as
$\chi_x \to e^{i\theta\epsilon_x}\chi_x$
where $\epsilon_x=(-1)^{x_0+\cdots+x_{d}}$ is related to 
$\gamma_5$ \cite{Susskind:1976jm,Sharatchandra:1981si,Kawamoto:1981hw}.

Throughout this paper, we consider the case of color $\mathit{SU}(N_c=3)$
in 3+1 dimension $(d=3)$ spacetime.
Temporal and spatial lattice sizes are denoted as $N_\tau$ and $L$,
respectively.
While $T=1/N_{\tau}$ takes discrete values, 
we consider $T$ as a continuous valued temperature.
We take account of finite $T$ effects
by imposing periodic and antiperiodic 
boundary conditions on link variables and quark fields, respectively.
We take the static and diagonalized gauge (called Polyakov gauge)
for temporal link variables with respect for the periodicity~\cite{DKS}.
In these setups, we derive the effective potential
$\Feff{eff}{}=-\log\bigl[{\cal Z}_{\mathrm{LQCD}}\bigr]/(N_{\tau}L^d)$
on the basis of the strong coupling expansion
with Polyakov loop effects.

\subsection{Effective action with Polyakov loop effects}\label{subsec:actionPL}
To treat the chiral and deconfinement phase transitions simultaneously,
we evaluate the leading order Polyakov loop effects
in the pure Yang-Mills sector
and NNLO effects in the fermionic sector.
In a finite $T$ treatment of SC-LQCD,
we first derive an effective action
from the spatial link ($U_j$) integrals,
and the temporal link ($U_0$) integral is evaluated later to consider
the thermal effects of quarks~\cite{Faldt:1985ec,DHK1985}.

\begin{figure}[th]
\Psfig{8cm}{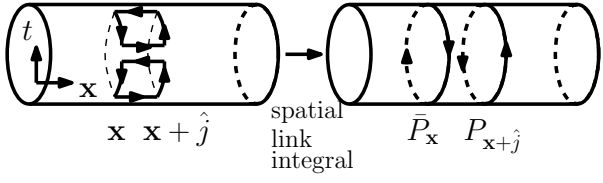}
\caption{
Leading order of the Polyakov loop effects in the strong coupling expansion.
The squares in the left and loops in the right 
represent the temporal plaquettes 
and the Polyakov loops, respectively.}
\label{Fig:Polyakov-loop-SC-LQCD}
\end{figure}

We consider the leading order Polyakov loop effective action $\Seff{P}$
of the strong coupling expansion in the pure Yang-Mills sector,
which is obtained from the spatial link integral
of the $N_\tau$ plaquettes sequentially connected in the temporal direction
as shown in Fig.~\ref{Fig:Polyakov-loop-SC-LQCD}.
\begin{align}
 \Seff{P}
=& -\left( \displaystyle \frac {1}{g^2 N_c} \right)^{N_{\tau}} N_c^2
     \sum_{\bfx, j>0} 
     \left(
      \bar{P}_\bfx P_{\bfx+ \hat{j}} + h.c.
     \right)
\ ,
\label{Eq:PLaction}
\end{align}
where $P_\bfx = \mathrm{tr}_c \prod_\tau U_0 (\tau,{\bf x}) / N_c$
represents the Polyakov loop. 
The factor $1/g^2N_c$ arises from the spatial link integral.
This effective action shows the nearest-neighbor interaction
of the Polyakov loop.
While we cannot reach the Stefan-Boltzmann limit 
in the high temperature limit with $\Seff{P}$~\cite{Langelage},
it is found to give results qualitatively consistent with the MC simulations
on the first order phase transition properties
such as the deconfinement temperature
and the discontinuity of the Polyakov loop
in the mean-field treatment of SC-LQCD~\cite{Kogut:1981ez,Fukushima:2003vi},
then essential dynamics around the deconfinement transition regime
may be described.

%============================================================
%============================================================

As for the effective action for quarks,
we utilize the NNLO effective action~\cite{Nakano:2009bf}
derived from one- and two-plaquette configurations,
which correspond to ${\cal O}(1/g^2)$ and ${\cal O}(1/g^4)$, 
respectively,
in addition to the strong coupling limit terms (${\cal O}(1/g^0)$).
We consider only the leading order terms of the $1/d$ expansion,
$\mathcal{O}(1/d^0)$,
which corresponds to taking minimum quark number configurations
for a given plaquette geometry~\cite{KlubergStern:1982bs}.
We briefly explain the NNLO effective action in Appendix~\ref{App:NNLO}.

The effective action including both of the quark and Polyakov loop
contributions is represented as
\begin{align}
S_\mathrm{eff}=& \Seff{F}+\Seff{X}+\Seff{P}
\ .\label{Eq:Seff}
\end{align}
$\Seff{F}$ represents the NNLO fermionic effective action, 
\begin{align}
S_\mathrm{eff}^{(F)}
=& Z_\chi \sum_{xy} \chibar_{x}\,
	G^{-1}_{xy}(\tilmq;\tilmu,T)\chi_{y}
\ ,
\label{Eq:SeffF}\\
G^{-1}_{xy}(\mq
&;\mu,T)
= \displaystyle \frac {1}{2} \left[
	 e^{\mu}U_{0,x}\,\delta_{x+\hat{0},y}
	-e^{-\mu}U^\dagger_{0,x}\,\delta_{x-\hat{0},y}
	\right]
\nn\\
&+ \mq \delta_{xy}
\ .
\end{align}
Finite coupling effects result in the modifications of 
the wave function renormalization factor $Z_\chi$,
quark mass $\tilmq$,
and effective chemical potential $\tilmu$.
$\Seff{X}$ represents the auxiliary field part of the effective action
described in the Appendix~\ref{App:NNLO}.

From a naive counting of the strong coupling expansion order,
the leading order Polyakov loop action composed of the plaquettes
is in the higher order ($\mathcal{O}(1/g^{2N_\tau})$), 
compared with the NNLO terms ($\mathcal{O}(1/g^{4})$)
stem from the quark sector.
As discussed in Introduction, 
we set the starting point of chiral and deconfinement dynamics
as the effective action of ${\cal O}(1/g^0)$ and ${\cal O}(1/g^{2N_\tau})$
in the quark and Polyakov loop sectors, respectively.
The effective action in Eq.~(\ref{Eq:Seff}) corresponds
to an extension in the quark sector
from this starting point.

From these action terms, 
an approximate QCD partition function is obtained,
and the effective potential 
is defined by the logarithm of the partition function as
\begin{align}
\Feff{eff}{}
\equiv& -\frac{1}{N_\tau L^d}\log\left[\int {\cal D}[\chi,\chibar,U_0]
	e^{-\Seff{F}-\Seff{X}-\Seff{P}}
	\right]
\nn\\
=&\Fq(\Phi;\mu,T)
 + U_g(\ell,\lbar)
 +\Feff{eff}{(X)}(\Phi)
\ .
\end{align}
Here $\Fq(\Phi;\mu,T),U_g(\ell,\lbar)$,
and $\Feff{eff}{(X)}(\Phi)$
represent the quark free energy, pure gluonic potential,
and the effective potential 
including only the auxiliary fields,
respectively.
We obtain $\Fq(\Phi;\mu,T)$ and $U_g(\ell,\lbar)$
by evaluating the Grassmann $(\chi,\bar{\chi})$
integral and the temporal link ($U_0$) integral,
\begin{align}
&\int {\cal D}U_0\, e^{-\Seff{P}} \mathrm{Det}\,
	\bigl[\Zchi G^{-1}(\tilde{m}_q;\tilde{\mu}, T)\bigr] 
\nn\\
&=\prod_{\mathbf{x}}
    e^{N_c(\log\Zchi+E_q)/T}
	\int d\mathcal{U}_0\, e^{-\Seff{P}}
\nn\\
& \times
	\mathrm{det}_c \Bigl[ 
	\left(1+\mathcal{U}_0 e^{-(E_q-\tilde{\mu})/T}\right)
	\left(1+\mathcal{U}_0^{\dagger} e^{-(E_q+\tilde{\mu})/T}\right)
	\Bigr]
\ , 
\label{Eq:eVq}
\end{align}
where $\mathcal{U}_0(\mathbf{x})=\prod_\tau U_0(\mathbf{x},\tau)$.
Note that the determinant is for the spacetime and color
in the first line of Eq. (\ref{Eq:eVq}),
and for the color in the second line of Eq. (\ref{Eq:eVq}). 
In Eq. (\ref{Eq:eVq}),
we carry out these integral as follows~\cite{DKS,Faldt:1985ec,Nishida:2003fb}.
First, we perform the Fourier transformation in the temporal direction,
and obtain the product in the frequency using the Grassmann integral
over the quark fields.
Second, we evaluate the product in the frequency by the Matsubara method.
Finally, in this paper, we evaluate the temporal link integral
in two kinds of schemes,
the Haar measure and Weiss mean-field methods.
We explain these methods in the next two subsections.

\subsection{Haar measure method}
\label{sec:Feffwo}

We shall now derive the effective potential with Polyakov loop effects
in the Haar measure method (H-method).
In the H-method, we replace the Polyakov loop with a mean-field value
and the Haar measure is taken into account in the Polyakov loop potential
instead of carrying out the temporal link integral.
The contribution to the effective action is,
\begin{align}
 \Seff{P}
\simeq& -2 \bp L^d \lbar \ell 
\ , \label{Eq:PLactionHaar}
\end{align}
where $\beta_p = (1/g^2 N_c)^{N_\tau} N_c^2 d$,
$\ell=\langle P_\bfx \rangle$ and $\lbar=\langle \bar{P}_\bfx \rangle$.
We assume the mean fields $\ell$ and $\lbar$
are constant and isotropic.

The temporal link integral is represented as the Haar measure
in the Polyakov gauge,
\begin{align}
\int d\mathcal{U}_0
=& \int d \ell d\lbar \cdot 27 
\left[ 1 -6 \ell \lbar +4 \left( \ell^3 + \lbar^3 \right) - 3 \left( \ell \lbar \right)^2 
\right]
\ , 
\end{align}
The Polyakov gauge is a static and diagonalized gauge 
for temporal link variables~\cite{DKS},
then ${\cal U}_0$ is given as follows, 
\begin{align}
{\cal U}_0(\bold{x})=\prod_\tau U_0(\bold{x},\tau)=\mathrm{diag}_c(e^{i\theta_1},e^{i\theta_2},e^{i\theta_3})
\ .
\end{align}
This Haar measure shows the Jacobian in the transformation
from the temporal link variables ($U_0$)
to the Polyakov loop ($\ell,\lbar$).
Therefore, $\Fq(\Phi;\mu,T)$ and $U_g(\ell,\lbar)$ are given as,
\begin{align}
&\Fq
=
- N_c E_q
- T \log R(E_q-\tilmu,N_c\ell, N_c\lbar)
\nn\\
&~~~~~~~~~- T \log R(E_q+\tilmu,N_c\lbar, N_c\ell)
- N_c \log \Zchi
\ ,\label{Eq:FqHaar}\\
&R(x,L,\bar{L})\equiv
1+Le^{-x/T}+\bar{L}e^{-2x/T}+e^{-3x/T}
\ , \\
&U_g
=
-2 T \bp \lbar \ell 
-T \log \left[
1 -6 \ell \lbar +4 \left( \ell^3 + \lbar^3 \right) - 3 \left( \ell \lbar \right)^2 
\right]
\ ,
\label{Eq:UgHaar}
\end{align}
where $\beta_p = (1/g^2 N_c)^{1/T} N_c^2 d$.
Here we have replaced the $N_\tau$ with $1/T$.
Note that we have omitted irrelevant constants.
In Eq. (\ref{Eq:FqHaar}), the first line represents the vacuum and quark 
free energy and the second line represents the antiquark 
free energy and the contribution of the wave function renormalization factor.
The quark free energy includes one- and two-quark excitations $(e^{-(E-\tilmu)/T},e^{-2(E-\tilmu)/T})$.
In the confined phase ($\ell \sim \lbar \sim 0$), 
the one- and two- quark excitations are suppressed and
only the color-singlet state contributions remain.
The pure gluonic potential $U_g$ does not include the fluctuation of the Polyakov loop since we 
treat the Polyakov loop as the mean field without the temporal link integral.
In the PNJL model, this pure gluonic potential is incorporated
to express the properties of the deconfinement phase transition~\cite{Fukushima:2003fw}.

%============================================================
\begin{figure}[tbh]
\Psfig{8cm}{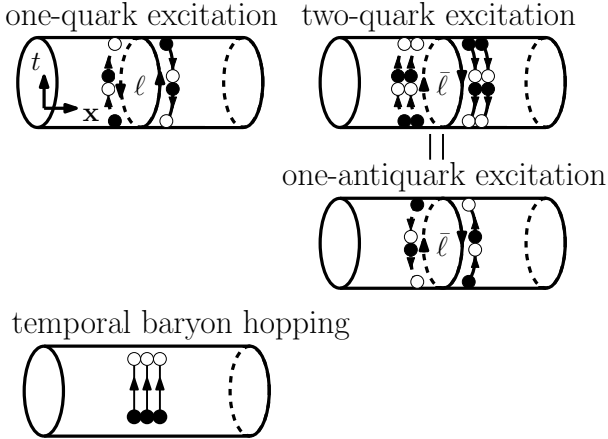}
\caption{Quark excitation on the lattice.
We show the one- and two-quark excitation in the first line, while the temporal baryon hopping
in the third line.
}
\label{Fig:Quark-excitation}
\end{figure}
%============================================================

\subsection{Weiss mean-field method}
\label{sec:FeffPL}
Now we shall evaluate the Polyakov loop effects
in the Weiss mean-field method (W-method).
The W-method 
includes some part of the fluctuation effects of the Polyakov loop.
We first bosonize the Polyakov loop action
by using the {\it Extended Hubbard-Stratonovich (EHS) transformation}~\cite{NLOSC-LQCD},
\begin{align}
e^{\alpha AB} \approx \exp\left[-\alpha(\bar{\psi}\psi-A\psi-\bar{\psi}B)\right]
\ ,
\end{align}
where $\alpha$ denotes a positive constant,
and two auxiliary fields are introduced simultaneously,
$(\bar{\psi}, \psi)=(\langle{A}\rangle,\langle{B}\rangle)$).
EHS transformation is a procedure to bosonize the product
of different types of composites~~\cite{NLOSC-LQCD}.
The Polyakov loop action in Eq.~(\ref{Eq:PLaction}) is linearized as,
\begin{align}
 \Seff{P}
\approx& \left( \displaystyle \frac {1}{g^2 N_c} \right)^{N_{\tau}} N_c^2  \sum_{{\bf x}, j>0} 2
     \left(
      \lbar \ell - \bar{P}_\bfx \ell - \lbar P_\bfx
     \right) \nonumber\\
\simeq&  2 \bp L^d \lbar \ell 
 -2 \beta_p \sum_{\bf x} \left( \bar{P}_\bfx \ell + \lbar P_\bfx \right)
\ ,
\label{Eq:PLactionEHS}
\end{align}
where $\bp,P_\bfx$, and $\bar{P}_\bfx$ are defined
in Subsec. \ref{subsec:actionPL}.
$\ell$ and $\lbar$ represent the auxiliary fields for the Polyakov loop,
$(\ell=\langle P_\bfx \rangle, \lbar=\langle \bar{P}_\bfx \rangle)$.
In the last line of Eq. (\ref{Eq:PLactionEHS}),
we assume constant and isotropic values for auxiliary fields $\ell$ and $\lbar$.

After the Grassmann integral,
we carry out the temporal link integral in Eq.~(\ref{Eq:eVq})
explicitly in the Polyakov gauge.
Since $\Seff{P}$ includes the Polyakov loop
$(P_\bfx,\bar{P}_\bfx)=(\mathrm{tr}\,{\cal U}_0/N_c, \mathrm{tr}\,{\cal U}_0^\dagger/N_c)$,
we have to include $\Seff{P}$ in the temporal link integral in the W-method.
The relevant part of the temporal link integral in Eq.~(\ref{Eq:eVq})
is given as,
\begin{align}
Z_P
=&\int d\mathcal{U}_0\, \mathrm{det}_c \Bigl[
	2\cosh(E_q/T)
	+\mathcal{U}_0 e^{\tilmu/T}
	+\mathcal{U}_0^{\dagger} e^{-\tilmu/T}
	\Bigr]
\nn\\
&\times \exp
\left[\eta\mathrm{tr}\,{\cal U}_0^\dagger+\bar{\eta}\mathrm{tr}\,{\cal U}_0\right]
\\
\equiv&e^{N_c E_q /T}Z_P^\prime
\ .
\label{Eq:eVqB}
\end{align}
We have defined $\eta=2\bp \ell /N_c$ and $\bar{\eta}=2\bp \lbar /N_c$.
In Appendix \ref{App:U0}, we explain the detail of the temporal link integral.
Eventually, we derive the effective potential,
\begin{align}
{\cal F}_q
= & -N_c E_q-T \log(Z_P^\prime/L_0) - N_c \log Z_\chi
\ ,\label{Eq:Fqboso}
\end{align}
\begin{subequations}
\begin{align}
Z_P^\prime/L_0
= &  
R(E_q-\tilmu,L_1,\Lbar_{1})
\,
R(E_q+\tilmu,\Lbar_{1},L_1)
\label{Eq:ZpD1a}
\\
+&
 e^{-2E_q/T} \Biggl[
\left( 1 + e^{-2E_q/T} \right)
\label{Eq:ZpD1b}
\\
+&
\left( 2 + L_2 - L_1 \Lbar_{1} \right)
\left( 1 + e^{-2E_q/T} \right)
\label{Eq:ZpD1c}
\\
+&
\left( 2 L_1 + L_3 - \Lbar_{1}^2 \right)
e^{ -(E - \tilmu ) /T} 
\label{Eq:ZpD1d}
\\
+&
\left( 2 \Lbar_{1} + \Lbar_{3} - L_1^2 \right)
e^{ -(E + \tilmu ) /T} 
\Biggr]
\label{Eq:ZpD1e}
\ ,
\end{align}
\end{subequations}
\begin{align}
U_g(\ell,\lbar)=& 2 T \bp \lbar \ell -T \log L_0
\label{Eq:Ugboso}
\ . 
\end{align}
$L_0, L_1, \Lbar_1, L_2, L_3, \Lbar_3$ are the functions of $\eta$ and $\etabar$.
These functions include the modified Bessel function $I_n(2\sqrt{\etabar \eta})$
and the ratio of $\ell$ and $\lbar$ 
$(\phi=\frac{1}{2}\log(\etabar/\eta)=\frac{1}{2}\log(\lbar/\ell))$.
The explicit expressions of them are shown in the Appendix~\ref{App:U0}.

\begin{figure}[tbh]
\Psfig{8cm}{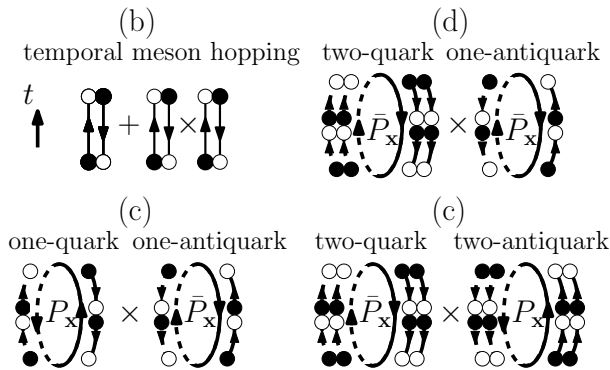}
\caption{(b)temporal meson hopping (c)one-quark and one-antiquark excitations, 
two-quark and two-antiquark excitations (d)one-antiquark and two-quark excitations}
\label{Fig:Quark-excitation2}
\end{figure}

Compared with the effective potential in the H-method
where the quark contributions are represented as the vacuum, quark and antiquark parts
, Eq. (\ref{Eq:FqHaar}),
the effective potential is more complicated.
While Eq.~(\ref{Eq:ZpD1a}) represents quark and antiquark parts 
(including the one- and two-quark excitations, and temporal baryon hopping) as H-method,
we replace $\ell(\lbar)$ with $P_\mathbf{x}(\Pbar_\mathbf{x})$ in Fig. \ref{Fig:Quark-excitation} .
The combination of the Bessel function accompany with the Boltzmann factors in W-method.
Other terms represent the temporal meson hopping (Eq.~(\ref{Eq:ZpD1b})), 
one- or two- quark and one- or two- antiquark excitations (Eq.~(\ref{Eq:ZpD1c})),
and one-antiquark (one-quark) and two-quark (two-antiquark) excitations 
(Eqs.~(\ref{Eq:ZpD1d}),(\ref{Eq:ZpD1e})), as shown 
in Fig. \ref{Fig:Quark-excitation2}.
We can understand these physical interpretations by considering the Boltzmann factors.
We explain the one-quark and one-antiquark excitations $(e^{-2E_q/T})$ in Eq.~(\ref{Eq:ZpD1c})
as an example.
We can decompose Eq.~(\ref{Eq:ZpD1a}), and find that the contribution of $(e^{-2E_q/T})$ comes
from the multiplication of one-quark $(e^{-(E_q-\tilmu)/T})$ and one-antiquark $(e^{-(E_q+\tilmu)/T})$
excitations.
These contributions of Eqs.~(\ref{Eq:ZpD1b})-(\ref{Eq:ZpD1e})
in W-method can appear because of the temporal link integral
and not in H-method or the PNJL model.
In the confined phase ($\ell, \lbar \sim 0$), 
the quark excitations are suppressed since
$L_1 \sim \Lbar_1 \sim L_3 \sim \Lbar_3 \sim \Lbar_2 + 2 \sim0$,
and the color-singlet state is dominant.
Therefore, Eqs.~(\ref{Eq:ZpD1a}) and (\ref{Eq:ZpD1b}) remain finite in the confined phase,
while the other terms Eq.~(\ref{Eq:ZpD1c})-(\ref{Eq:ZpD1e}) vanish.

Note that the pure gluonic potential $U_g$ includes the dependence
on $\lbar/\ell$ explicitly (i.e. the dependence on $\mu$).
When the quark chemical potential is zero ($\mu=0$),
the Polyakov loop for quarks and antiquarks are the same $(\ell=\lbar)$.
In comparison,
when the quark chemical potential is finite ($\mu \neq 0$),
the Polyakov loop for antiquarks
is generally different from that for quarks
$(\ell \neq \lbar)$~\cite{llbarldiff}.  

In Ref. \cite{Kogut:1981ez}, the pure gluonic potential
is derived at $\mu=0$ without quarks.
As shown later, we confirmed that our results are consistent with their results,
for example, in the temperature of the deconfinement phase transition without quarks.

%============================================================
%============================================================

\section{CHIRAL AND DECONFINEMENT PHASE TRANSITION}\label{sec:chiraldeconf}

We shall now discuss the phase transitions in P-SC-LQCD.
We impose the stationary condition for the effective potential
with respect to the auxiliary fields,
and obtain thermodynamical quantities.
In this article, we investigate the chiral and deconfinement
transitions at $N_c=3$ and $\mu=0$.
First, we discuss the deconfinement transition in pure Yang-Mills theory,
and next we show the results in the W-method.
Finally we compare the results in the W- and H-methods.

\subsection{Deconfinement transition in pure Yang-Mills theory}
%%%%%%%%%%%%%%%%%%%%%%%%%%%%%%%%%%%%%%%%%%%%%%%%%%%%%%%%%%%%%%%%%%%%%%%%%%%%%%%%
\begin{figure}[tbh]
\Psfig{8cm}{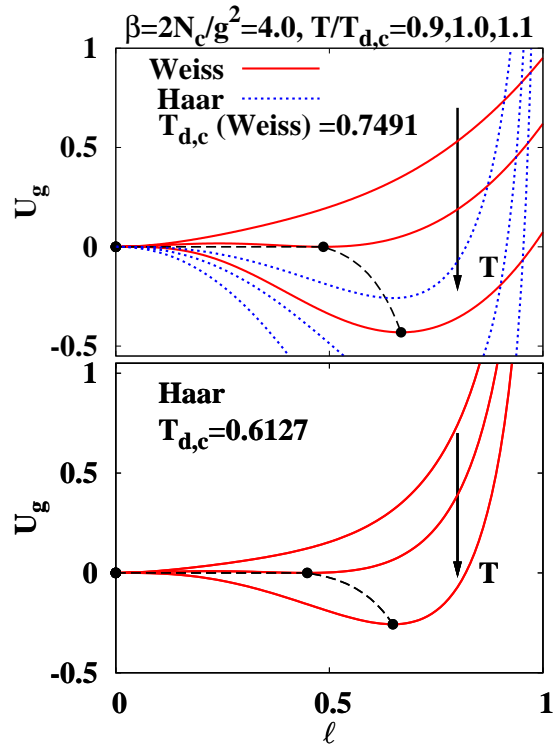}
\caption{Polyakov loop potential in the 
W- (upper panel) and  
H- (lower panel) methods.
The filled circles show the equilibrium points.
We show the results in the lattice unit.
}
\label{Fig:PandP_H-Polyakov-loop-potential}
\end{figure}
%%%%%%%%%%%%%%%%%%%%%%%%%%%%%%%%%%%%%%%%%%%%%%%%%%%%%%%%%%%%%%%%%%%%%%%%%%%%%%%%

Before discussing the results with quarks,
we first examine
the deconfinement transition in the pure Yang-Mills theory.
In Fig. \ref{Fig:PandP_H-Polyakov-loop-potential},
we compare the Polyakov loop potentials $U_g$ in the W- and H-methods,
Eqs. (\ref{Eq:Ugboso}) and (\ref{Eq:UgHaar}), respectively.
We show the results at $T=(0.9-1.1) T_{d,c}$ at $\beta=4$
without quark effects.
Both of the potentials have qualitatively the same behavior
and show the first-order transitions at $T=T_{d,c}$.
We find some quantitative differences in these two potentials.
First, we note that in H-method there is a wall at $\ell=1$ where $U_g$ diverges.
The range of $\ell$ is limited in H-method to be less than one as can be found
from Eq. (\ref{Eq:UgHaar}).
In the W-method,
the integral over ${\cal U}_0$ smoothens the H-method potential,
\begin{align}
e^{-U^W_g(\ell,\lbar)/T}=\int d\ell' d\lbar' e^{-\beta_P (\ell-\ell')(\lbar-\lbar')}
	e^{-U^H_g(\ell',\lbar')/T}
\ ,
\label{Eq:W-H-Ug}
\end{align}
and the effective potential becomes a milder function of $\ell$.
As shown in the dotted line in the upper panel of Fig.~\ref{Fig:PandP_H-Polyakov-loop-potential},
$U_g^W$ is smaller (larger) for large (small) $\ell$.
Since the transition temperature is more sensitive to the potential at small $\ell$,
the $T_{d,c}$ in the W-method is higher than that in the H-method.

%=========================================================================
\begin{table}[t]
\caption{Correspondence of the parameters between our results and Ref.~\cite{Kogut:1981ez}.
Note that we replace $\beta$ in Ref.~\cite{Kogut:1981ez}
with $\beta^{(K)}$ to distinguish from $\beta=2N_c/g^2$.
}
\label{Tab:Corrrespondance-Kogut}
\begin{center}
\begin{tabular}{cc}
\hline\hline
 Our results & Kogut et al.~\cite{Kogut:1981ez} $(d=3)$\\
\hline
$\ell$ & $\alpha /2d$ \\
$2 \beta_p / N_c^2$ & $2d \beta^{(K)} / (2 N_c)$ \\
\hline\hline
\end{tabular}
\end{center}
\end{table}
%=========================================================================

Our treatment in W-method in the pure Yang-Mills theory
is the same as the chiral spin model investigated in Ref.~\cite{Kogut:1981ez}.
where the following effective action is adopted,
\begin{align}
\Seff{K}/L^d=\frac{\beta^\mathrm{(K)}N_c}{4d}\alpha^2
-\frac{\beta^\mathrm{(K)}N_c}{2}\alpha\left(P+\bar{P}\right)
\ .
\end{align}
When the parameter $\beta^\mathrm{(K)}$ and the variable $\alpha$
are rewritten according to Table \ref{Tab:Corrrespondance-Kogut},
this action is found to be equivalent to $\Seff{P}$
in Eq.~(\ref{Eq:PLactionEHS}).
In Ref. \cite{Kogut:1981ez}, the first-order phase transition is found
to take place at the critical coupling $\beta^\mathrm{(K)}_c=0.81$
in the mean-field treatment.
By using the relation $\beta^\mathrm{(K)}=2\beta_p/N_cd$,
this critical coupling corresponds to $T_{d,c} \sim 0.75$ at $\beta=4$.
This value is consistent with our result
$T_{d,c} \sim 0.7491$ in the upper panel of
Fig.~\ref{Fig:PandP_H-Polyakov-loop-potential}.

\subsection{Chiral condensate and Polyakov loop}\label{subsec:Feffresults}

\begin{figure}
\Psfig{8cm}{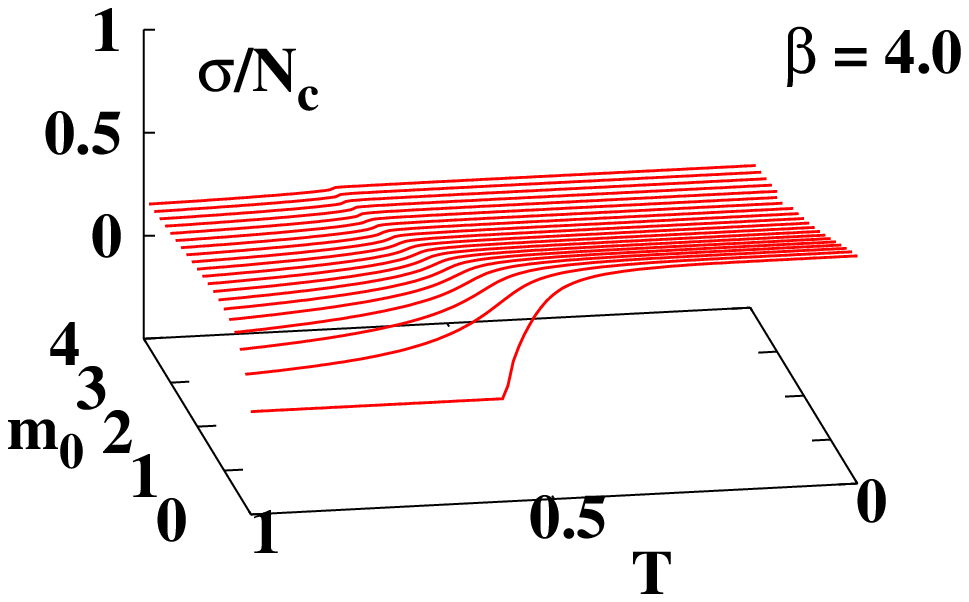}
\Psfig{8cm}{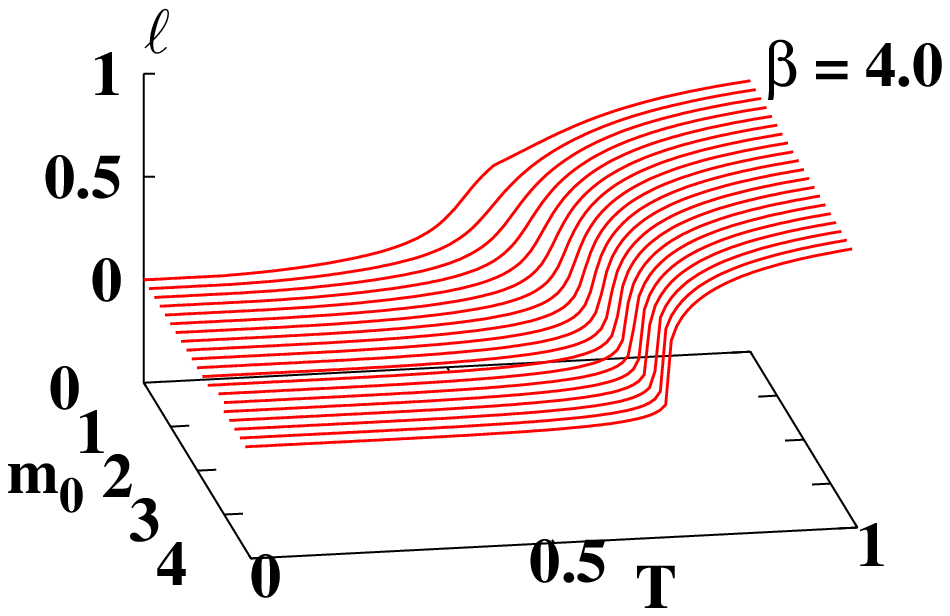}
\caption{Temperature and quark mass dependence 
of the chiral condensate (upper panel)
and the Polyakov loop (lower panel) in the W-method.
We show both of the results in the lattice unit.
}\label{Fig:TMdep}
\end{figure}

\begin{figure}
\Psfig{8cm}{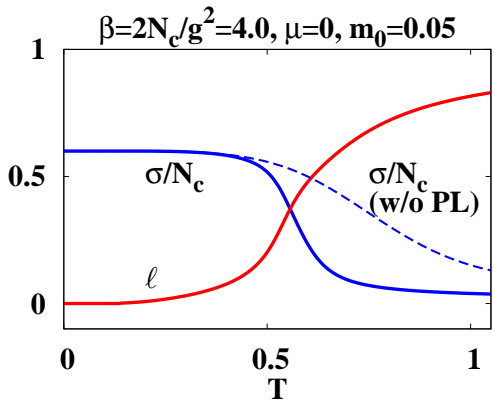}
\caption{Chiral condensate and Polyakov loop 
in P-SC-LQCD (solid lines),
and chiral condensate in SC-LQCD without the Polyakov loop effects
(dashed line) as functions of $T$ at $\mu=0$ in the W-method.
We show the results in the lattice unit.
}
\label{Fig:P-mu0-sig-Ploop-NNLO-A}
\end{figure}

\begin{figure}[tbh]
\Psfig{8cm}{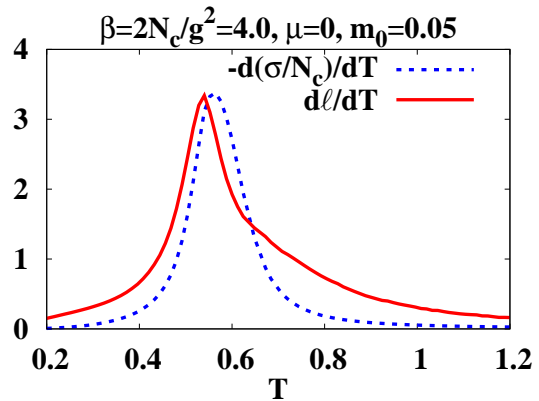}\\
\Psfig{8cm}{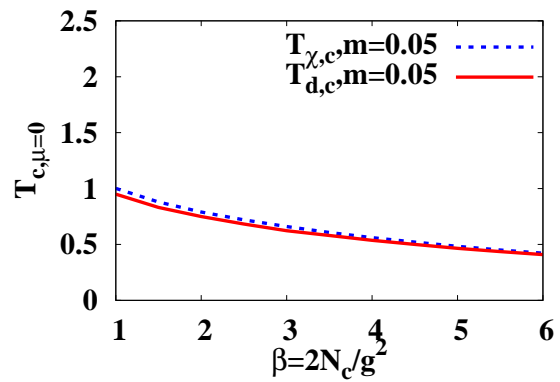}
\caption{
Upper panel: 
Temperature dependence of
$-d\sigma/dT$ (dashed line) and $d \ell /dT$ (solid line).
%Middle panel: 
%Temperature dependence of
%the susceptibility $\chi_\sigma$ (dashed line) and $\chi_\ell$ (solid line).
Lower panel: 
$\beta$ dependence of the critical temperatures for the chiral (dashed line) 
and deconfinement (solid line) phase transitions.
All of them are obtained in the W-method.
We show both of the results in the lattice unit.
}
\label{Fig:P-mu0-susceptibility-NNLO-A}
\end{figure}

\begin{figure*}[tbh]
\Psfig{10cm}{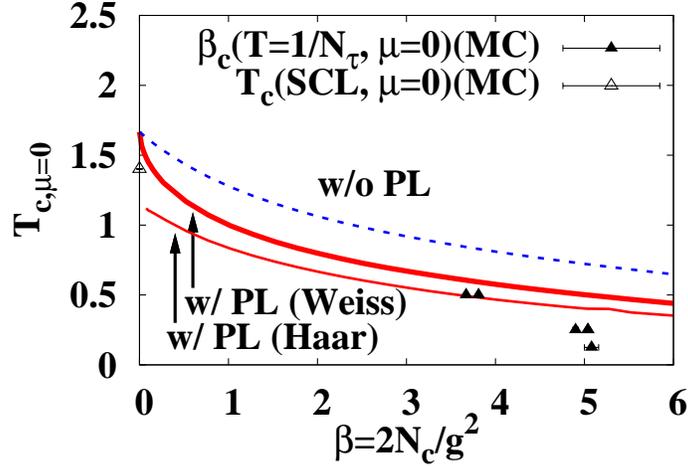}
\caption{
Comparison of P-SC-LQCD (solid line) 
and SC-LQCD without the Polyakov loop (dashed line) in the lattice unit.
We show the results of Weiss mean-filed method (bold solid line) and Haar measure method (thin solid line) in P-SC-LQCD at $m_0=0.0$. 
The triangles represent the results of
the critical temperature ($T_{c,\mu=0}$, open triangle)
and the critical coupling ($\beta_c$, filled triangles)
obtained in Monte-Carlo simulations
with one species of unrooted staggered fermion: 
From the left,
$T_{c,\mu=0}$ in the SCL with monomer-dimer-polymer (MDP) simulations~\cite{deForcrand:2009dh},
$\beta_c$ at $(N_\tau,m_0)=(2,0.025)$ \cite{Forcrand},
$(2,0.05)$ \cite{Forcrand},
$(4,0.0)$ \cite{Gottlieb:1987eg},
$(4,0.05)$ \cite{D'Elia:2002gd,Fodor:2001au},
and $(8,0.0)$ \cite{Gavai:1990ny}.
}
\label{Fig:Tc-W-H}
\end{figure*}

We shall now discuss the chiral and deconfinement transitions
in the W-method at finite $T$ and $\mu=0$,
via the $T$ dependence of the chiral condensate $\sigma = -\VEV{\chibar\chi}$
and the Polyakov loop
$(\ell, \lbar)=(\VEV{P_\mathbf{x}}, \VEV{\Pbar_\mathbf{x}})$.
In Fig.~\ref{Fig:TMdep}, we show the quark-mass ($m_0$) and $T$ dependence
of $\sigma$ and $\ell$ at $\mu=0$.
We adopt the coupling $\beta=4$ as a typical example.
As expected,
$\sigma$ vanishes sharply at $T_c$ at $m_0=0$
showing the second order transition in the chiral limit,
and $\ell$ starts to show the first order transition at $m_0 \sim 4.0$.

In Fig.~\ref{Fig:P-mu0-sig-Ploop-NNLO-A},
we show $\sigma$ and $\ell$ as functions of $T$
at $\beta=4$ and $m_0=0.05$.
Since the quark mass is small but finite,
both of $\sigma$ and $\ell$ show crossover transition
towards the chiral restored ($\sigma \to 0$)
and deconfined ($\ell \to 1$) matter,
i.e. a quark-gluon plasma.

From Fig.~\ref{Fig:P-mu0-sig-Ploop-NNLO-A},
the chiral and deconfinement transitions seem to take place 
almost at the same temperatures.
In the upper panel of Fig. \ref{Fig:P-mu0-susceptibility-NNLO-A},
we show $-d \sigma/dT$ and $d \ell/dT$ as functions of $T$.
Both of $-d \sigma/dT$ and $d \ell/dT$ are peaked at around $T \sim 0.55$.
We here define the chiral $T_{\chi,c}$ and deconfinement $T_{d,c}$ 
transition temperatures as the peak temperatures of $-d \sigma/dT$ and $d \ell/dT$.
In the lower panel of Fig. \ref{Fig:P-mu0-susceptibility-NNLO-A},
we display $T_{\chi,c}$ and $T_{d,c}$ as functions of $\beta$.

%=========================================================================
\begin{table}[t]
\caption{Difference between the SC-LQCD with/without the Polyakov loop (PL) effects, the PNJL,
and NJL model.}
\label{Tab:Diff}
\begin{center}
\begin{tabular}{ccc}
\hline\hline
 & low $T$ & high $T$ \\
\hline\hline
SC-LQCD w/o PL & \multicolumn{2}{c}{confined} \\
NJL & \multicolumn{2}{c}{deconfined} \\
P-SC-LQCD and PNJL& confined      & deconfined \\
\hline\hline
\end{tabular}
\end{center}
\end{table}
%=========================================================================

Polyakov loop is found to suppress the chiral condensate
and to reduce the chiral transition temperature in P-SC-LQCD.
In Figs.~\ref{Fig:P-mu0-sig-Ploop-NNLO-A}
and \ref{Fig:Tc-W-H},
we compare the chiral condensate $\sigma$
and the chiral transition temperature $T_{\chi,c}$
with and without Polyakov loop effects.
We find that both $\sigma$ and $T_{\chi,c}$ in P-SC-LQCD
are smaller than those without Polyakov loop effects;
the chiral condensate becomes smaller when the Polyakov loop
takes a finite value,
and the chiral symmetry restoration takes place at lower $T$
by the Polyakov loop effects.
In SC-LQCD without Polyakov loop effects,
we find the contribution only from 
color-singlet states as discussed before,
then we implicitly assume that the quarks are confined at any $T$.
In the W-method,
we have one- and two-quark contribution as well
when the Polyakov loop takes a finite value.
Quark excitation generally breaks the chiral condensates,
then it promotes the chiral symmetry to be restored at lower $T$.
The similar behavior is found also in the H-method.

This behavior
is different from that in the PNJL model \cite{Fukushima:2003fw,Ratti:2005jh}.
In PNJL, the Polyakov loop pushes up 
the chiral transition temperature from that in the NJL model. 
This difference can be understood as follows.
Quarks are deconfined in NJL, then the role of the Polyakov loop
is to {\em confine quarks at low $T$ in PNJL}.
By comparison,
only color-singlet hadronic states contribute to the effective potential
and quarks are confined in SC-LQCD w/o PL,
then the role of the Polyakov loop is to 
{\em deconfine quarks at high $T$ in P-SC-LQCD}.
As a result, confined and deconfined phases appear at low and high $T$ region
in both of PNJL and P-SC-LQCD,
and the upward and downward shifts of $T_c$ in PNJL and P-SC-LQCD
are understood as the directions to enlarge the region
which is promoted by the Polyakov loop.
As a result, both of PNJL and P-SC-LQCD give a consistent picture of the chiral and deconfinement
transitions, although the confinement properties of quark matter are different in NJL 
and SC-LQCD without Polyakov loop effects as shown in Table \ref{Tab:Diff}.

Note that the confinement picture of quark matter in PNJL model and P-SC-LQCD is not
completely the same.
In P-SC-LQCD, colors of quarks in a hadron are confined
locally when Polyakov loop are zero.
By comparison the inter-quark distance in a hadron
is not necessarily small in PNJL,
since the momentum of quarks are specified.

\begin{figure}[tbh]
\Psfig{7cm}{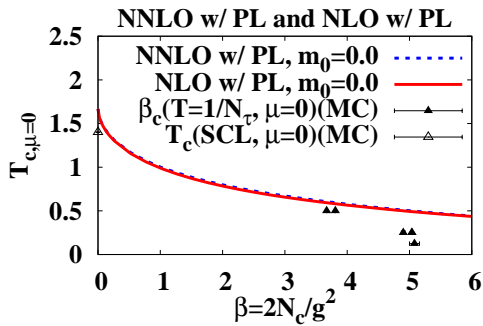}
\caption{
Comparison of the chiral transition temperature $T_{\chi,c}$
in P-SC-LQCD with NNLO (solid line) and NLO (dashed line)
finite coupling corrections for quarks.
The triangles represent
the same MC results in Fig. \ref{Fig:Tc-W-H}.
We show the results in the lattice unit.
}
\label{Fig:Tc-W-method}
\end{figure}

In our previous work on NLO~\cite{NLOSC-LQCD}
and NNLO~\cite{Nakano:2009bf} SC-LQCD, 
the critical temperature is calculated to be larger than MC results,
and NNLO effects on $T_{\chi,c}$ are found to be small.
In Figs. \ref{Fig:Tc-W-H} and \ref{Fig:Tc-W-method},
we show $T_{\chi,c}$ in several treatments of P-SC-LQCD
in comparison with the MC results.
As already discussed, 
the chiral transition temperature $T_{\chi,c}$
is reduced by the Polyakov loop effects
and roughly explains the MC results,
especially in the region $\beta \lesssim 4$.
By comparison, the NNLO effects on $T_{\chi,c}$ at $\mu=0$ are small
also in the case with Polyakov loop effects.
This observation implies that introducing the Polyakov loop,
the deconfinement order parameter, is essential to explain 
the QCD phase transition temperature.
In order to explain further reduction of $T_{\chi,c}$ at $\beta \gtrsim 4$,
we may need to include higher order terms of $1/g^2$
in the Polyakov loop effective action.

\subsection{Comparison of Weiss mean-field and Haar measure methods}
\label{subsec:Haar}
The Haar measure method is useful to include the confinement effects
in chiral effective models, and it has been widely applied
for example in the PNJL~\cite{Fukushima:2003fw} and PQM~\cite{Schaefer:2009ui} models.
In this subsection,
we discuss the results of the Haar measure method (H-method),
and compare the results of two methods.
\begin{figure}[tbh]
\Psfig{8cm}{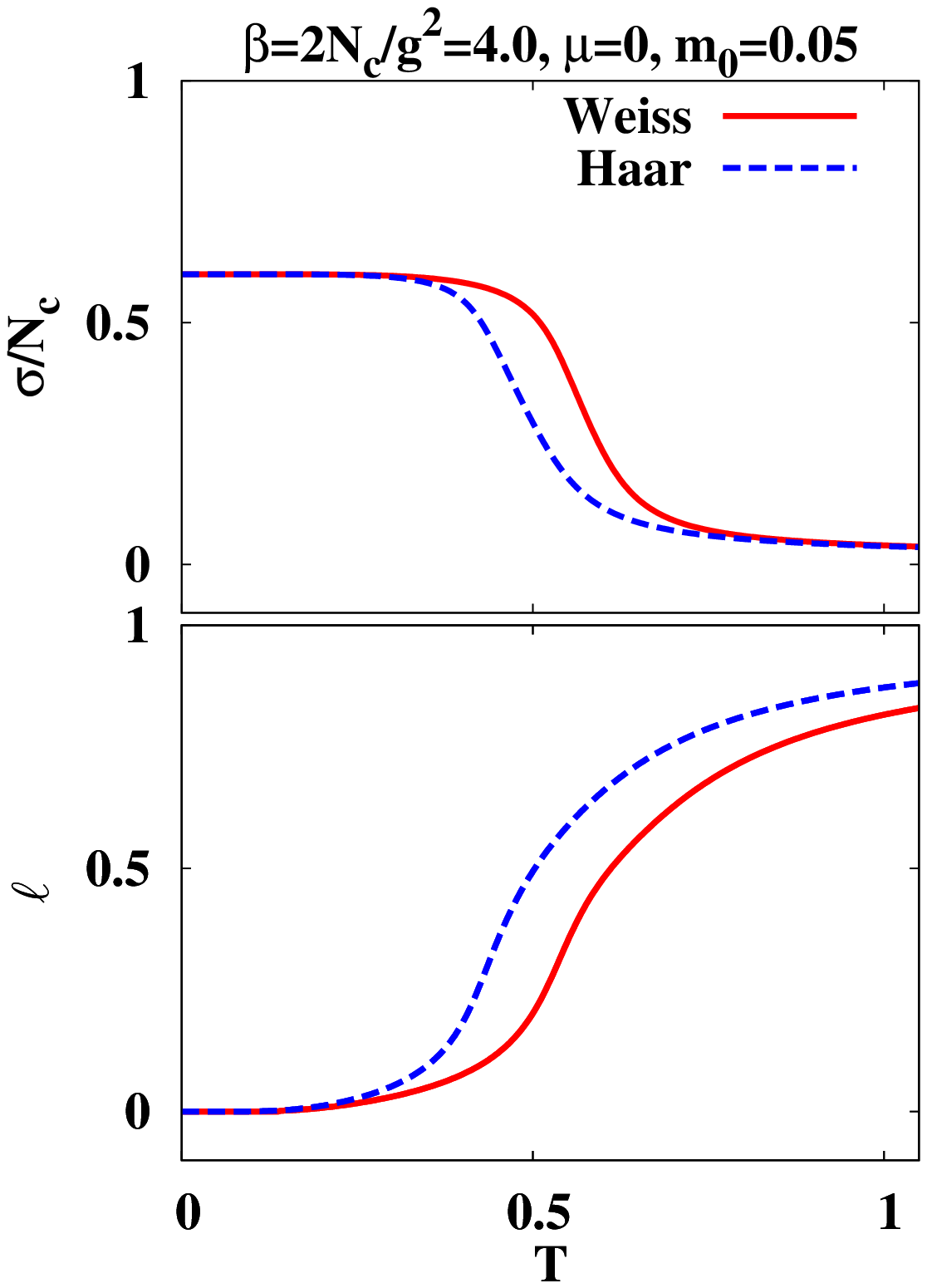}
\caption{Chiral condensate and Polyakov loop as functions of $T$ at $\mu=0$
in the W- and H-methods.
The notation is the same as Fig. \ref{Fig:P-mu0-sig-Ploop-NNLO-A}.
We show the results in the lattice unit.
}
\label{Fig:sigP-H-method}
\end{figure}
%
%%%%%%%%%%%%%%%%%%%%%%%%%%%%%%%%%%%%%%%%%%%%%%%%%%%%%%%%%%%%%%%%%%%%%%%%%%%%%%%%
\begin{figure}[tbh]
\Psfig{8cm}{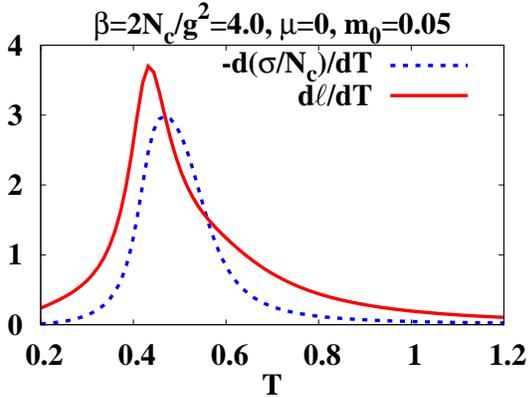}%\\
\caption{
Temperature dependence of 
$-d\sigma/dT$ (dashed line) and $d \ell /dT$ (solid line)
in the Haar measure method (H-method).
The notation is the same as Fig. \ref{Fig:P-mu0-susceptibility-NNLO-A}.
We show the results in the lattice unit.
}
\label{Fig:Tc-H-method}
\end{figure}
%%%%%%%%%%%%%%%%%%%%%%%%%%%%%%%%%%%%%%%%%%%%%%%%%%%%%%%%%%%%%%%%%%%%%%%%%%%%%%%%
%
%
%%%%%%%%%%%%%%%%%%%%%%%%%%%%%%%%%%%%%%%%%%%%%%%%%%%%%%%%%%%%%%%%%%%%%%%%%%%%%%%%
\begin{figure}[tbh]
\Psfig{8cm}{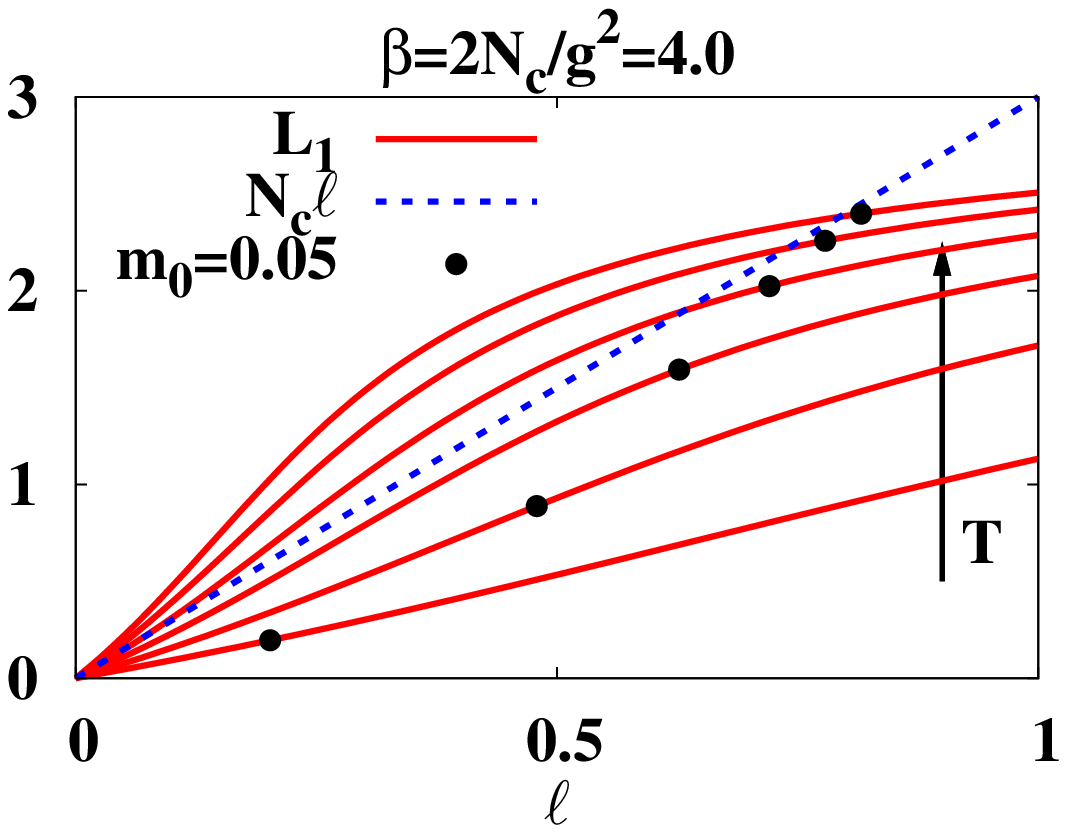}\\
\Psfig{8cm}{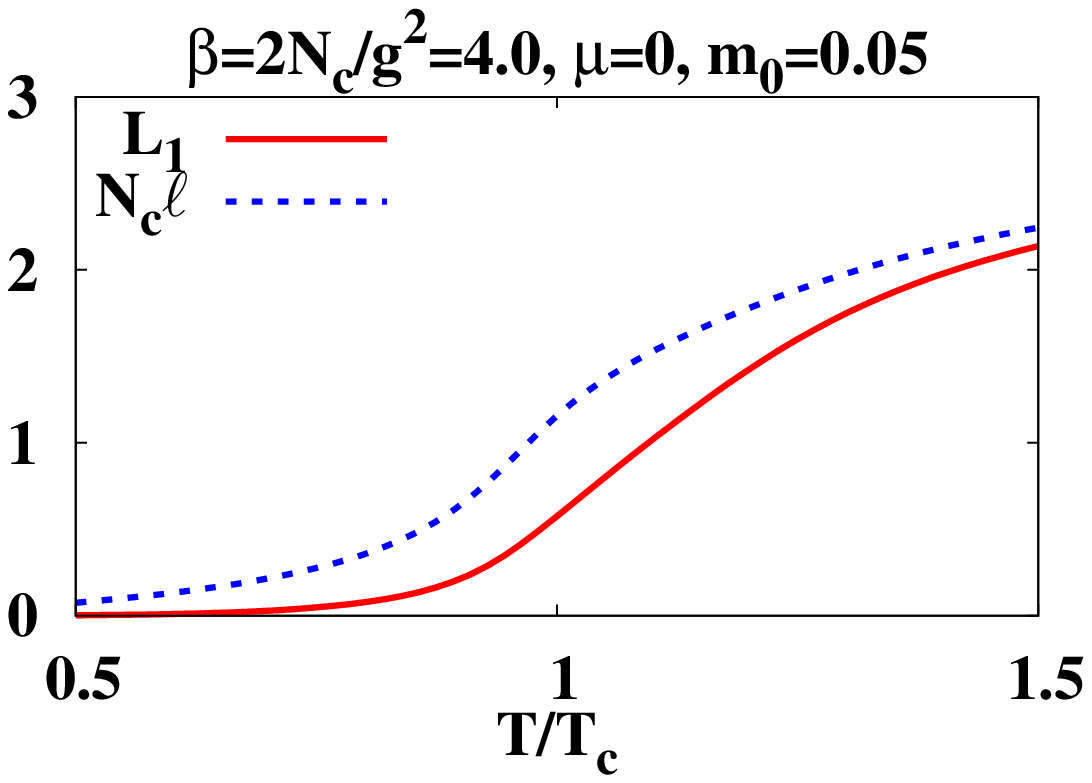}
\caption{
Upper panel: The Polyakov-quark coupling $L_1$ and $N_c\ell$
as functions of $\ell$ at $\mu=0$ (T=0.5,..,1.0) (solid line).
The dots represent the value of $L_1$
under the stationary condition ($m_0=0.05$). 
Lower panel: Stationary value of the Polyakov-quark coupling 
$L_1$ and $N_c\ell$
as functions of the normalized temperature $T/T_c$.
We show both of the results in the lattice unit.
}
\label{Fig:P-D-NNLO-A}
\end{figure}
%%%%%%%%%%%%%%%%%%%%%%%%%%%%%%%%%%%%%%%%%%%%%%%%%%%%%%%%%%%%%%%%%%%%%%%%%%%%%%%%

First we compare the chiral condensate and the Polyakov loop in the W- and H-methods.
In Fig. \ref{Fig:sigP-H-method},
we show the $T$ dependence of $\sigma$ and $\ell$ in the two methods.
While qualitative behaviors are the same in both of the methods,
the chiral condensate (Polyakov loop) in the W-method is larger (smaller)
than those in the H-method,
suggesting that the W-method exhibits a little larger $T_{c}$ than the H-method.
In Fig. \ref{Fig:Tc-W-H},
we compare the chiral transition temperature $T_{\chi,c}$ in the two methods.
The transition temperature is defined from 
the peaks of $-d\sigma/dT$ and $d \ell /dT$ 
%the susceptibility peak
as shown in Fig.~\ref{Fig:Tc-H-method}.

Physically, the temporal link integral in the Weiss mean-field method favors
the color-singlet states and therefore suppress the quark excitation.
Also, as another aspect,
the smaller Polyakov-quark coupling in the W-method
is responsible in part for the larger $T_c$.
In the W-method, we carry out the temporal link integral
of the effective action in which fermions couple
with the temporal link variables.
This "averaging" procedure as shown Eq.~(\ref{Eq:W-H-Ug}) smears the Polyakov-quark coupling.
In the upper panel of Fig.~\ref{Fig:P-D-NNLO-A},
we show the comparison of the Polyakov-quark couplings,
$L_1(=\Lbar_1)$ and $N_c\ell$ in the W- and H-methods, respectively,
defined as the factor accompanied with the Boltzmann factor.
At $\mu=0$, we find that the stationary value of coupling 
is smaller in the W-method, $L_1(=\Lbar_1) < N_c\ell$,
at any temperature as shown in Fig.~\ref{Fig:P-D-NNLO-A}.
These observations confirm the smaller Polyakov-quark coupling.
Similar behavior is also found in the NJL model
with the Polyakov integration~\cite{Megias:2004hj}.

\section{Concluding Remarks}\label{sec:CR}
In this paper,
we have derived an analytic expression of the effective potential
at finite temperature and chemical potential
in the strong coupling lattice QCD with the Polyakov loop effects
using one species of unrooted staggered fermion.
We have discussed the chiral and deconfinement transitions at $\mu=0$,
especially the Polyakov loop effects on these transitions.
We have considered NNLO effects in the strong coupling expansion
and the leading order of $1/d$ expansion in the quark effective action.
We have adopted the leading order %effects
Polyakov loop action,
${\cal O}(1/g^{2N_\tau})$,
and the temporal link ($U_0$) integral is evaluated in two methods.
One is the Haar measure method (H-method),
where we replace the Polyakov loop
with its constant mean-field without the $U_0$ integral.
The deconfinement dynamics is taken account of via
the Haar measure.
Another is the Weiss mean-field method (W-method),
where we bosonize the effective action using the Extended Hubbard-Stratonovich
transformation and carry out the temporal link integral explicitly.
The fluctuation effects appear
as the combination of the modified Bessel function
and the difference of $\lbar$ and $\ell$
in the effective potential.
We have found that one- and two-quark excitations are allowed in both methods,
when the Polyakov loop takes a finite value at high $T$.

We find that the Polyakov loop reduces
the chiral transition temperature,
and the obtained transition temperatures roughly explain
the MC results in the region $\beta \lesssim 4$.
The chiral and deconfinement transitions are found to take place
at similar temperatures, but the two transition temperatures are slightly
different when we define them from the susceptibilities.
We have compared the results in the W- and H-methods.
These methods exhibit qualitatively the same results,
but the Polyakov loop effects are found to be weaker in the W-method.
The transition temperature is slightly higher
and the Polyakov-quark coupling is smaller
in the W-method.
This is because the temporal link integral in the W-method
favors the color-singlet states
and leads to the suppression of the quark excitation.

There are several points to be studied further.
Firstly, we should investigate the QCD phase diagram in finite $T$ and $\mu$ 
on the basis of the effective potentials derived in this work.
Secondly, it would be a promising direction to extend
the present P-SC-LQCD framework
by including higher order terms in $1/g^2$ and $1/d$
in the quark and Polyakov loop effective action.
Thirdly, it is interesting to apply the present effective potential 
to the chiral effective model, for example to the PNJL model,
in order to include the fluctuation of the Polyakov loop.
We replace the Polyakov loop potential with that in Eq. (\ref{Eq:Ugboso})
and the factor accompanied with the Boltzmann factor
$N_c \ell \ (N_c \lbar )$ with $L_1 \ (\Lbar_1)$ in Eq. (\ref{Eq:Fqboso}).
These studies will be helpful to shed light on the QCD phase transitions
at finite $T$ and $\mu$.

\section*{Acknowledgments}
We would like to thank Hideo Suganuma and Yoshimasa Hidaka for useful discussion and advice.
This work was supported in part
by Grants-in-Aid for Scientific Research from MEXT and JSPS
(Nos. 22-3314),
the Yukawa International Program for Quark-hadron Sciences (YIPQS),
and by Grants-in-Aid for the global COE program
`The Next Generation of Physics, Spun from Universality and Emergence'
from MEXT.

\appendix
\section{NNLO effective action}
\label{App:NNLO}

We here briefly introduce the NNLO SC-LQCD
effective action~\cite{Nakano:2009bf},
which we utilize as the quark part of the P-SC-LQCD.
The NNLO effective action is
derived from one- and two-plaquette configurations,
which correspond to ${\cal O}(1/g^2)$ and ${\cal O}(1/g^4)$, 
respectively,
in addition to the strong coupling limit terms (${\cal O}(1/g^0)$).
We consider only the leading order terms of the $1/d$ expansion,
$\mathcal{O}(1/d^0)$.
The spatial link integral leaves a sum over spatial directions,
and the energy per bond is assumed to be proportional to $1/d$
in order to keep the action finite at large spatial dimension 
in the $1/d$ expansion \cite{KlubergStern:1982bs}.
Since the quark field scales as $d^{-1/4}$,
the leading order $1/d$ terms correspond to the minimum quark number
configurations for a given plaquette geometry.
The NNLO effective action is represented as~\cite{Nakano:2009bf}
\begin{align}
S_\mathrm{eff}^\mathrm{(NNLO)}=& \Seff{F}+\Seff{X}
\ ,\label{Eq:SeffNNLO}
\end{align}
$\Seff{F}$ represents the fermionic effective action, 
shown in Eq.~(\ref{Eq:SeffF}).
\begin{align}
S_\mathrm{eff}^{(F)}
=& \frac{1}{2}\sum_x\left[ \Zm V^+_x - \Zp V^-_x \right]
 + \sum_x m_q^{\prime} M_x 
\nn\\
=& Z_\chi \left[
	\sum_{x,y}
	  \frac12\left[
	  	 e^{-\delta\mu}V^+_x
		-e^{\delta\mu}V^-_x
	  \right]
	+\sum_x
	  \tilmq M_x
	\right]
\nn\\
=& Z_\chi \sum_{xy} \chibar_{x}\,
	G^{-1}_{xy}(\tilmq;\tilmu,T)\chi_{y}
\ ,
\label{Eq:SeffFapp}
\\
V_x^+ =& \bar{\chi}_x e^{\mu} U_{0,x} \chi_{x + \hat{0}}
\ ,\quad
V_x^- = \chibar_{x+\hat{0}} e^{-\mu} U_{0,x}^{\dagger} \chi_x
\ ,\\
M_x =& \chibar_x \chi_x
\ .
\end{align}
Finite coupling effects result in the modifications of 
the wave function renormalization factor
$Z_\chi=\sqrt{\Zp \Zm}$,
quark mass
$\tilmq=m_q^{\prime}/Z_\chi$,
and effective chemical potential
$\tilmu=\mu-\delta \mu=\mu-\log\sqrt{\Zp / \Zm}$.
$Z_{\pm}$ and $\mq^{\prime}$ are defined as,
\begin{align}
m_q^{\prime}=& \bsigp \sigma + m_0 - \btt(\psibarTT+\psiTT)
\ ,\\
\Zp=& 1+\bt'\psibarT+4\btt m_q^{\prime}\psibarTT
\ ,\\
\Zm=& 1+\bt'\psiT   +4\btt m_q^{\prime}\psiTT
\ .
\end{align}
$\Seff{X}$ represents the 
auxiliary field part of the effective action,
\begin{align}
\Feff{eff}{(X)}
=&\Seff{X}/(N_\tau L^d)
\nn\\
=&\frac{1}{2} \bsigp \sigma^2
+\frac{1}{2}\bt'\bar{\psi}_\tau\psi_\tau
+\frac{1}{2}\beta_s'\psibarS\psiS
\nn\\
+&\beta_{\tau\tau}\psibarTT\psiTT
+\beta_{ss}\bar{\psi}_{ss}\psi_{ss}
+\frac{1}{2}\beta_{\tau s}\bar{\psi}_{\tau s}\psi_{\tau s}
\label{Eq:SeffX}
\ ,
\end{align}
Auxiliary fields $(\sigma,\{ \psi_K, \bar{\psi}_K ; K= \tau, s, \tau\tau, \tau s, ss \})$
in Eq. (\ref{Eq:SeffX}) are introduced as mean fields
for the mesonic composites.
We show the relation between auxiliary fields and the mesonic elements in Table \ref{Tab:aux}. 
The coefficients of the auxiliary field 
in Eq. (\ref{Eq:SeffX}) are defined as
\begin{align}
&\bsigp=\bsig +
2 \left[ \beta_{ss}\psi_{ss}
+\beta_{\tau s}\bar{\psi}_{\tau s}
+\beta_s'(\psiS+\psibarS)
\right]
\ ,\\
&\bt'=\bt+\beta_{\tau s}\psi_{\tau s}
\ ,\quad
\beta_s'=\beta_s+2\beta_{ss}\bar{\psi}_{ss}
\ ,\\
&b_\sigma=\frac{d}{2N_c}\ ,\quad
\bt=\frac{d}{N_c^2g^2}\,\left(1+\frac{1}{2g^2}\right)
\ ,\\
&\beta_s=\frac{d(d-1)}{8N_c^4g^2}\,\left(1+\frac{1}{2g^2}\right)
\ ,\quad
\beta_{\tau\tau}=\frac{d}{2N_c^3g^4}
\ ,\\
&\beta_{ss}=\frac{d(d-1)(d-2)}{16N_c^7g^4}
\ ,\quad
\beta_{\tau s}=\frac{d(d-1)}{2N_c^5g^4}
\ .\label{Eq:CoeffB}
\end{align}

%=========================================================================
\begin{table}[t]
\caption{The auxiliary fields in the NNLO effective potential. 
$W_{\bfx}^+$ and $W_{\bfx}^-$ are represented as 
the mesonic composite which connects 
the quark fields in the next-to-nearest neighboring temporal
sites ($W_{\bfx}^+=\bar{\chi}_x e^{2\mu} U_{0,x} U_{0,x+\hat{0}} \chi_{x+2\hat{0}}$ and 
$W_{\bfx}^-=\bar{\chi}_{x+2\hat{0}} e^{-2\mu} U_{0,x+\hat{0}}^{\dagger} U_{0,x}^{\dagger} 
\chi_{x}$). 
We omit the indices $x$ ($\mathbf{x}$ in the case of $\ell,\lbar$) 
in the mesonic composites.
}\label{Tab:aux}
\begin{center}
\begin{tabular}{cc}
\hline\hline
Auxiliary Fields	&Mean Fields		\\
\hline
$\sigma$		&$ -M $			\\
$(\psibarTT,\psiTT)$	&$(W^+,W^-)$		\\
$(\psibarSS,\psiSS)$	&$(MM,MMMM)$		\\
$(\psibarTS,\psiTS)$	&$(-\Vp \Vm,2MM)$	\\
$(\psibarT,\psiT)$	&$(-\Vp, \Vm)$		\\
$(\psibarS,\psiS)$	&$(MM, MM)$		\\
$(\lbar,\ell)$		&$(\Pbar, P)$		\\
\hline\hline
\end{tabular}
\end{center}
\end{table}
%=========================================================================

\section{The temporal link integral in the Weiss mean-field method}
\label{App:U0}
The part of the temporal link integral in Eq. (\ref{Eq:eVq}) is given as,
\begin{align}
&Z_{P}=\int d\mathcal{U}_0\,
\exp \left[ 2 \bp ( \Pbar_\bfx \ell + \lbar P_\bfx )  \right]
\,e^{N_cE_q/T}
\nn\\
&\times\mathrm{det}_c \left\{
	\left[1+\mathcal{U}_0 e^{-(E_q-\tilmu)/T}\right]
	\left[1+\mathcal{U}_0^{\dagger} e^{-(E_q+\tilmu)/T}\right]
	\right\}
\nn\\
=&\int [d\mathcal{U}_0] 
\prod_{a=1}^{N_c} 
\exp \left( \eta e^{-i \theta_a} + \etabar e^{i \theta_a} \right)
\left[X+Ye^{i\theta_a}+\frac{e^{-i\theta_a}}{Y}\right]
\ .
\end{align}
We define $\eta=2\bp \ell /N_c$, $\etabar=2\bp \lbar /N_c$,
$X=2\cosh(N_\tau E_q)$
and
$Y=\exp(N_\tau \tilmu)$.

This type of temporal link integral is written for $N_c=3$
as~\cite{Nishida:2003fb},
\begin{align}
&\int [d\mathcal{U}_0] 
     \prod_{a=1}^{N_c} 
     f(\theta_a)    
= \sum_{n=-\infty}^{\infty} N_c! \det_{i,j} M_{n+i-j}
\nn\\
=& N_c! \sum_n \left[M_n^3+M_{n+1}^2M_{n-2}+M_{n-1}^2M_{n+2}
\right.
\nn\\
&
~~~~~~~~~
-M_n(M_{n+2}M_{n-2}+2M_{n+1}M_{n-1})
\left. \right]
\ ,
\label{Eq:Zp}
\end{align}
where
\begin{align}
M_{n}    
&=\int_{-\pi}^{\pi} \displaystyle \frac{d\theta}{2\pi} f(\theta) 
\exp\left[-i n \theta \right] 
\ , 
\label{Eq:Ml}\\
f(\theta) &= e^{iz\sin(\theta-\pi/2-i\phi)} 
\left[X+Ye^{i\theta}+\frac{e^{-i\theta}}{Y}\right]\ ,
\end{align}
where
$z=2i\sqrt{\eta\etabar}$
and 
$\phi=\frac12\log(\etabar/\eta)$.
By using to the relation
$e^{iz\sin\theta}= \sum_{l=-\infty}^{\infty} J_l(z) e^{il\theta}$,
where $J_l(z)$ is the Bessel function
and $l, z$ are an integer and a complex number, respectively,
$f(\theta)$ is found to be,
\begin{align}
f(\theta) =& 
\sum_{l=-\infty}^{\infty} J_l(z) e^{il(\theta-\pi/2-i\phi)} 
\left[X+Ye^{i\theta}+\frac{e^{-i\theta}}{Y}\right]\ ,
\label{Eq:ftheta}
\end{align}
We substitute Eq. (\ref{Eq:ftheta}) into Eq. (\ref{Eq:Ml}) and
carry out the integral over $\theta$.
The matrix element $M_{n}$ is calculated to be
\begin{align}
M_{n}
=& e^{n\phi}\left[I_{n}(y) X
+ I_{n-1}(y) e^{-\phi} Y
+ I_{n+1}(y) \frac{e^{\phi}}{Y}
\right]
\label{Eq:Mn}
\ .
\end{align}
Note that $I_\nu(y)=e^{-\nu \pi i/2} J_{\nu}(iy)$ is
the modified Bessel function, where 
$y=2\sqrt{\eta\etabar}$.
By substituting Eq.~(\ref{Eq:Mn}) into Eq.~(\ref{Eq:Zp}),
the temporal integral is found to be (up to a constant ($N_c!$)),
\begin{align}
Z_P=&D_1 X^3+D_2 X+D_3\tilY^3+D_{-3}\tilY^{-3}
\nn\\
+&X^2(D_4\tilY+D_{-4}\tilY^{-1})
+X(D_5\tilY^2+D_{-5}\tilY^{-2})
\nn\\
+&D_6\tilY+D_{-6}\tilY^{-1}
\ ,
\end{align}
where $\tilY=Ye^{-\phi}$.
Coefficients $D_{i} \ (i=1,2,\pm3, \cdots,\pm6)$ are defined as, 
\begin{align}
D_{i}(\eta,\etabar) &\equiv \sum_{n=-\infty}^{\infty} e^{N_c n \phi} D_{n,i}
\ , \label{Eq:Dsum}\\
%------------------
D_{n,1}
&= I_n^{3}-I_{n+2}I_nI_{n-2}
 - 2 I_{n+1}I_nI_{n-1}
\nn\\
&+ I_{n+1}^2I_{n-2}
+ I_{n-1}^2I_{n+2} 
\ , \label{Eq:D1}\\
%------------------
D_{n,2}
&=
-2(I_n^{3}-I_{n+2}I_nI_{n-2})
+5 I_{n+1}I_nI_{n-1}
\nn\\
&-3(I_{n+1}^2I_{n-2}+I_{n-1}^2I_{n+2}) 
- I_{n+3}I_nI_{n-3}
\nn\\
&+ I_{n-1}I_{n-2}I_{n+3} + I_{n+1}I_{n+2}I_{n-3}
\ , \label{Eq:D2}\\
%------------------
D_{n,\pm3}
&=D_{n\mp1,1}
\ , \label{Eq:D3}\\
%------------------
D_{n,\pm4}
&=
I_n^2I_{n\mp1}
+I_{n\pm1}^2I_{n\mp3} 
\nn\\
&
-I_{n\mp1}^2I_{n\pm1}
+I_{n\pm2}I_{n\mp1}I_{n\mp2} 
-I_{n\pm2}I_{n}I_{n\mp3}
\nn\\
&
- I_{n\pm1}I_nI_{n\mp2} 
\ , \\
%------------------
D_{n,\pm5}
&=D_{n\mp1,\mp4}
\ , \label{Eq:D5}\\
%------------------
D_{n,\pm6}
&=
-(I_{n}^2I_{n\mp1}+I_{n\pm1}^2I_{n\mp3})
\nn\\
&
+2(I_{n\mp1}^2I_{n\pm1}
-I_{n\pm2}I_{n\mp1}I_{n\mp2} 
+I_{n\pm2}I_nI_{n\mp3})
\nn\\
&
+I_{n\mp2}^2I_{n\pm3}
-I_{n\pm3}I_{n\mp1}I_{n\mp3} 
\ .
\end{align}
In the numerical calculation,
we have confirmed that the sums in $D_k$ converge.
$D_{n,i}$ $(i=1,2,\pm 3, \cdots, \pm 6)$
are represented as combinations of the modified Bessel functions. 
From Eqs.~(\ref{Eq:D3}) and (\ref{Eq:D5}),
$D_{\pm3}$ and $D_{\pm5}$ are represented as $D_{\pm3}=e^{\pm N_c \phi}D_1$ 
and $D_{\pm5}=e^{\pm N_c \phi}D_{\mp4}$.
We find $D_{i^\prime}(\eta,\etabar)=D_i(\etabar,\eta)$
from the relation $D_{-n,i^\prime}=D_{n,i}$
($i^\prime=i$ for $i=1,2$ and $i^\prime=-i$ for $i=3,\cdots , 6$).
By utilizing these properties, 
$Z_P$ is rewritten as
\begin{align}
Z_P/D_1=&X^3+Y^3+Y^{-3}+\tilD{2} X
\nn\\
+&X^2(\tilD{4}Y+\tilD{-4}Y^{-1})
+X(\tilD{-4}Y^2+\tilD{4}Y^{-2})
\nn\\
+&\tilD{6}Y+\tilD{-6}Y^{-1}
\ ,\\
\tilD{2}\equiv D_2/&D_1 
\ ,\quad
\tilD{\pm k}=D_{\pm k}e^{\mp\phi}/D_1\  (k=4,6)
\ .
\end{align}

When the quarks are confined ($\ell \to 0, \lbar \to 0$),
the 0th modified Bessel function remains finite ($I_0 \to 1$)
but others vanish, $I_n(y) \to 0 (n>0, n<0)$.
In this limit, $Z_P$ 
reproduces 
the standard SC-LQCD expression without the Polyakov loop
effects,
\begin{align}
Z_P \to& X^3-2X+Y^3+Y^{-3}
\nn\\
=& \frac{\sinh((N_c+1)E_q/T)}{\sinh(E_q/T)}+2\cosh(N_c\tilmu/T)
\ ,\\
D_1 \to & 1
\ ,\quad
D_2 \to -2
\ .
\end{align}
In this main part, we redefine $L_0 \equiv D_1, L_1 \equiv \tilD{4}, \Lbar_1 \equiv \tilD{-4}, 
L_2 \equiv \tilD{2},
L_3 \equiv \tilD{6}, \Lbar_3 \equiv \tilD{-6}$.

%%%%%%%%%%%%%%%%%%%%%%%%%%%%%%%%%%%%%%%%%%%%%
%%%%%%%%%%%%%%%%%%%%%%%%%%%%%%%%%%%%%%

%%%%%%%%%%%%%%%%%%%%%%%%%%%%%%%%%%%%%%
\end{document}